\documentstyle[11pt]{article}
\makeatletter
\def\revtex@ver{1.6}
\def\revtex@date{12 Aug 93}
\def\revtex@org{PASP}
\def\revtex@jnl{}
\def\revtex@genre{conference proceedings}
\def\revtex@pageid{\xdef\@thefnmark{\null}
\@footnotetext{This \revtex@genre\space was prepared with the
\revtex@org\space \revtex@jnl\space Rev\TeX\ macros v\revtex@ver.}}
\ifnum\@ptsize<1
\fi
\def\ps@paspcstitle{\let\@mkboth\@gobbletwo
\def\@oddhead{\null{\footnotesize\it\@slug}\hfil}
\def\@oddfoot{\rm\hfil\thepage\hfil}
\let\@evenhead\@oddhead\let\@evenfoot\@oddfoot
}
\def\ps@myheadings{\let\@mkboth\@gobbletwo
\def\@oddhead{\hbox{}\hfil\sl\rightmark\hskip 1in\rm\thepage}%
\def\@oddfoot{}%
\def\@evenhead{\rm\thepage\hskip 1in\sl\leftmark\hfil\hbox{}}%
\def\@evenfoot{}\def\sectionmark##1{}\def\subsectionmark##1{}}
\def\@leftmark#1#2{\sec@upcase{#1}}
\def\@rightmark#1#2{\sec@upcase{#2}}
\def\@singleleading{0.9}
\def\@doubleleading{1.6}
\def\baselinestretch{\@singleleading}
\def\tightenlines{\def\baselinestretch{\@singleleading}}
\def\loosenlines{\def\baselinestretch{\@doubleleading}}
\clubpenalty\@M
\widowpenalty\@M
\def\@journalname{ASP Conference Series}
\def\cpr@holder{Astronomical Society of the Pacific}
\def\@jourvol{10000}
\def\cpr@year{1994}
\def\vol@title{Astronomical Data Analysis Software and Systems III}
\def\vol@author{R.\ J.\ Hanisch, D.\ R.\ Crabtree, and J.\ Barnes, eds.}
\let\journalid=\@gobbletwo
\let\articleid=\@gobbletwo
\let\received=\@gobble
\let\accepted=\@gobble
\def\@slug{{\tabcolsep\z@\begin{tabular}[t]{l}\vol@title\\
\@journalname, Vol.\ \@jourvol, \cpr@year\\
\vol@author
\end{tabular}}
}
\def\paspconf@frontindent{.45in}
\def\title#1{\vspace*{1.0\baselineskip}
\@tempdima\textwidth \advance\@tempdima by-\paspconf@frontindent
\hfill
\parbox{\@tempdima}
	{\pretolerance=10000\raggedright\large\bf\sec@upcase{#1}}\par
\vspace*{1\baselineskip}\thispagestyle{title}}
\def\author#1{\vspace*{1\baselineskip}
\@tempdima\textwidth \advance\@tempdima by-\paspconf@frontindent
\hfill
\parbox{\@tempdima}
{\pretolerance=10000\raggedright{#1}}\par}
\def\affil#1{\vspace*{.5\baselineskip}
\@tempdima\textwidth \advance\@tempdima by-\paspconf@frontindent
\hfill
\parbox{\@tempdima}
{\pretolerance=10000\raggedright{\it #1}}\par}

\def\abstract{\vspace*{1.3\baselineskip}\bgroup\leftskip\paspconf@frontindent
\noindent{\bf\sec@upcase{Abstract.}}\hskip 1em}
\def\endabstract{\par\egroup\vspace*{1.4\baselineskip}}
\skip\footins 4ex plus 1ex minus .5ex
\long\def\@makefntext#1{\noindent\hbox to\z@{\hss$^{\@thefnmark}$}#1}

\def\tablenotetext#1#2{
\@temptokena={\vspace{.5ex}{\noindent\llap{$^{#1}$}#2}\par}
\@temptokenb=\expandafter{\tblnote@list}
\xdef\tblnote@list{\the\@temptokenb\the\@temptokena}}
\def\spewtablenotes{
\ifx\tblnote@list\@empty
\else
\let\@temptokena=\tblnote@list
\gdef\tblnote@list{\@empty}
\vspace{4.5ex}
\footnoterule
\vspace{.5ex}
{\footnotesize\@temptokena}
\fi}
\newtoks\@temptokenb
\def\tblnote@list{}
\def\endtable{\spewtablenotes\end@float}
\@namedef{endtable*}{\spewtablenotes\end@dblfloat}
\let\tableline=\hline
\def\thefigure{\@arabic\c@figure}
\def\fnum@figure{Figure \thefigure.}
\def\thetable{\@arabic\c@table}
\def\fnum@table{Table \thetable.}
\long\def\@makecaption#1#2{
\vskip 10pt
\setbox\@tempboxa\hbox{#1\hskip 1.5em #2}
\let\@tempdima=\hsize \advance\@tempdima by -2em
\ifdim \wd\@tempboxa >\@tempdima
	{\leftskip 2em
	#1\hskip 1.5em #2\par}
\else
	\hbox to\hsize{\hskip 2em\box\@tempboxa\hfil}
\fi}
\def\fps@figure{tbp}
\def\fps@table{htbp}
\let\keywords=\@gobble
\let\subjectheadings=\@gobble
\def\upper{\def\sec@upcase##1{\uppercase{##1}}}
\def\sec@upcase#1{\relax#1}
\def\section{\@startsection {section}{1}{\z@}{-4.2ex plus -1ex minus
-.2ex}{2.2ex plus .2ex}{\normalsize\bf}}
\def\subsection{\@startsection{subsection}{2}{\z@}{-2.2ex plus -1ex minus
-.2ex}{1.1ex plus .2ex}{\normalsize\bf}}

\def\subsubsection{\@startsection{subsubsection}{3}{\z@}{-2.2ex plus
-1ex minus -.2ex}{-1.2em}{\normalsize\it}}
\def\thesection{\@arabic\c@section.}
\def\thesubsection{\thesection\@arabic\c@subsection.}
\def\thesubsubsection{\thesubsection\@arabic\c@subsubsection.}
\def\@sect#1#2#3#4#5#6[#7]#8{\ifnum #2>\c@secnumdepth
\def\@svsec{}\else
\refstepcounter{#1}\edef\@svsec{\csname the#1\endcsname\hskip 1em }\fi
\@tempskipa #5\relax
\ifdim \@tempskipa>\z@
\begingroup #6\relax
\@hangfrom{\hskip #3\relax\@svsec}{\interlinepenalty \@M \sec@upcase{#8}\par}%
\endgroup
\csname #1mark\endcsname{#7}\addcontentsline
{toc}{#1}{\ifnum #2>\c@secnumdepth \else
\protect\numberline{\csname the#1\endcsname}\fi
#7}\else
\def\@svsechd{#6\hskip #3\@svsec #8\csname #1mark\endcsname
{#7}\addcontentsline
{toc}{#1}{\ifnum #2>\c@secnumdepth \else
\protect\numberline{\csname the#1\endcsname}\fi
#7}}\fi
\@xsect{#5}}
\def\@ssect#1#2#3#4#5{\@tempskipa #3\relax
\ifdim \@tempskipa>\z@
\begingroup #4\@hangfrom{\hskip #1}{\interlinepenalty \@M \sec@upcase{#5}\par}%
\endgroup
\else \def\@svsechd{#4\hskip #1\relax #5}\fi
\@xsect{#3}}
\def\acknowledgments{\@startsection{paragraph}{4}{1em}
{1ex plus .5ex minus .5ex}{-1em}{\bf}{\sec@upcase{Acknowledgments.}}}

\def\qanda@heading{Discussion}
\newif\if@firstquestion \@firstquestiontrue
\newenvironment{question}[1]{\if@firstquestion
\section*{\qanda@heading}\global\@firstquestionfalse\fi
\par\vskip 1ex
\noindent{\it#1\/}:}{\par}
\newenvironment{answer}[1]{\par\vskip 1ex
\noindent{\it#1\/}:}{\par}
\def\mathwithsecnums{
\@newctr{equation}[section]
\def\theequation{\hbox{\normalsize\arabic{section}-\arabic{equation}}}}
\def\references{\section*{References}
\bgroup\parindent=0pt\parskip=.5ex
\def\refpar{\par\hangindent=3em\hangafter=1}}
\def\endreferences{\refpar\egroup}

\def\@biblabel#1{\relax}
\def\@cite#1#2{#1\if@tempswa , #2\fi}
\def\reference{\relax\refpar}
\def\@citex[#1]#2{\if@filesw\immediate\write\@auxout{\string\citation{#2}}\fi
\def\@citea{}\@cite{\@for\@citeb:=#2\do
{\@citea\def\@citea{,\penalty\@m\ }\@ifundefined
{b@\@citeb}{\@warning
{Citation `\@citeb' on page \thepage \space undefined}}%
{\csname b@\@citeb\endcsname}}}{#1}}
\let\jnl@style=\rm
\def\ref@jnl#1{{\jnl@style#1\/}}
\def\aj{\ref@jnl{AJ}}
\def\araa{\ref@jnl{ARA\&A}}
\def\apj{\ref@jnl{ApJ}}
\def\apjl{\ref@jnl{ApJ}}
\def\apjs{\ref@jnl{ApJS}}
\def\ao{\ref@jnl{Appl.Optics}}
\def\apss{\ref@jnl{Ap\&SS}}
\def\aap{\ref@jnl{A\&A}}
\def\aapr{\ref@jnl{A\&A~Rev.}}
\def\aaps{\ref@jnl{A\&AS}}
\def\azh{\ref@jnl{AZh}}
\def\baas{\ref@jnl{BAAS}}
\def\jrasc{\ref@jnl{JRASC}}
\def\memras{\ref@jnl{MmRAS}}
\def\mnras{\ref@jnl{MNRAS}}
\def\pra{\ref@jnl{Phys.Rev.A}}
\def\prb{\ref@jnl{Phys.Rev.B}}
\def\prc{\ref@jnl{Phys.Rev.C}}
\def\prd{\ref@jnl{Phys.Rev.D}}
\def\prl{\ref@jnl{Phys.Rev.Lett}}
\def\pasp{\ref@jnl{PASP}}
\def\pasj{\ref@jnl{PASJ}}
\def\qjras{\ref@jnl{QJRAS}}
\def\skytel{\ref@jnl{S\&T}}
\def\solphys{\ref@jnl{Solar~Phys.}}
\def\sovast{\ref@jnl{Soviet~Ast.}}
\def\ssr{\ref@jnl{Space~Sci.Rev.}}
\def\zap{\ref@jnl{ZAp}}

\def\la{\mathrel{\hbox{\rlap{\hbox{\lower4pt\hbox{$\sim$}}}\hbox{$<$}}}}
\def\ga{\mathrel{\hbox{\rlap{\hbox{\lower4pt\hbox{$\sim$}}}\hbox{$>$}}}}

\newcount\lecurrentfam
\def\LaTeX{\lecurrentfam=\the\fam \leavevmode L\raise.42ex
\hbox{$\fam\lecurrentfam\scriptstyle\kern-.3em A$}\kern-.15em\TeX}

\newif\if@finalstyle \@finalstylefalse
\if@finalstyle
\ps@myheadings
\let\ps@title=\ps@paspcstitle
\else
\ps@plain
\let\ps@title=\ps@plain
\fi
\ds@twoside
\makeatother
\begin{document}
\thispagestyle
\markright{\tiny\noindent To appear in {\bf Barred Galaxies}, IAU
Coll.~157, R.~Buta, B.G.~Elmegreen \& D.A.~Crocker
(eds.), ASP Series, (1996)}

\title{Stellar Dynamics and the 3D Structure of Bars}
\author{D. Pfenniger}
\affil{Geneva Observatory, CH-1290 Sauverny, Switzerland}
\begin{abstract}
Recent observational constraints restrict the strict applicability of stellar
dynamics in spirals to a few rotation periods.  However, stellar dynamics
concepts such as periodic orbits are invaluable for understanding the various
dynamical processes occurring during much more periods.  A distinction of two
instability types in stellar systems is pointed out, the first one being well
illustrated by the bar instability, and the second one by the bar bending
instability.  In bars the third dimension brings essential dynamical effects
which modify the views about the history of bulges and the spiral secular
evolution.  Bars may grow, bend, thicken, and dissolve into spheroidal bulges,
and spirals may evolve along the Hubble sequence in the sense Sd$\to$Sa.  This
leads to a much more dynamical picture of isolated galaxies than imagined
before.
\end{abstract}


\section{Applicability of Stellar Dynamics in Spirals}
Recent developments about spirals bring a quite different picture about their
physical state.  The discovery of their large far-infrared (FIR) flux by IRAS
and COBE, comparable or sometimes superior to the optical one, and consistent
with the evidences that their optical parts are semi-transparent, means that a
substantial fraction of the stellar light is thermalized, and re-emitted by
dust in the FIR.  This is {\it not negligible\/} with respect to the the
typical
power that large-scale dynamical processes (spiral arms, bars) can exchange,
the gravitational power (ratio of the gravitational energy to the dynamical
time).  For a system in near virial equilibrium, $v^2 = GM/R$, we have $L_{\rm
grav}\approx\left(GM^2/R\right)\big/\left(R/v\right)=v^5/G$.  Replacing $v$ by
the typical rotation speeds of spirals we find powers surprisingly close to
the the galaxy luminosities (e.g.~$10^{44}\,\rm erg\, s^{-1}$ for the Milky
Way).  Since the IR Tully-Fisher relation ($L_{\rm H} \propto v^{4-5}$) is
nearly parallel to this equation, the match is close along the whole spiral
sequence (for more references on the subject see Pfenniger 1992).

This coincidence supports the proposition that a feedback mechanism relates
both dynamical instabilities and stellar activity.  Essential is that although
a stellar population pours out energy over several Gyr, the first Myr of a
starburst is a quicker cooperative and intense reaction with respect to the
dynamical time $\tau_{\rm dyn}$.  The idea of a feedback mechanism between
dynamics and star formation has been proposed several times (e.g.~Quirk 1972;
Kennicutt 1989).  With a light thermalization power of the order of the
gravitational power, we just need to convert at an intermediate stage (between
light production and thermalization) a sizable fraction of this power into
{\it mechanical energy}.  It is quite obvious in images of HII regions and
starbursts that a substantial mass of surrounding gas is set into coherent
motion by the massive star heating; this yields ultimately HI holes and
superbubbles.  If a fraction $\gamma$ of the stellar power is converted into
mechanical power, the time-scale to change significantly the global binding
energy of the whole galaxy is $\tau_{\rm dyn}/\gamma$.  If
$0.02\!<\!\gamma\!<\!1$ then the evolution time-scale associated with stellar
activity is shorter than the galaxy age; the galaxy as a whole must depend on
the stellar energy output. $\gamma$ is not well known but is estimated to be
in the range $\gamma\approx 0.1$ in starbursts (Leitherer \& Heckman 1995).

Such simple facts change fundamentally the way to see spirals.  In the early
days of galactic dynamics, galaxies were viewed as essentially transparent
collections of stars, therefore the stellar energy output could be discounted,
exactly as the huge supernova neutrino flux.  Also the large scale dynamical
instabilities such as bars were not understood, only the slow 2-body
relaxation was considered as a factor of evolution, then the dimension and
shape of galaxies had to be viewed as determined by the initial conditions of
formation.  The subsequent evolution was viewed as so slow that stellar
dynamics concepts could be used for several Gyr.  This led to absorb all the
dynamical effects in a rigid potential in many models, such as the stellar
population synthesis ones.

The conjunction of new elements on the energetics of spirals forces us to
modify the way to use stellar dynamics.  Which of the concepts such as
relaxation, integral of motion, etc., are still relevant and what are their
new limits of applicability?  If we take the Milky Way as a template of the
spirals, from observational data it is clear that the virial {\it gross
equilibrium}, at least in the optical disk, must be consistent with a
dominance of the bulk kinetic energy $E_{\rm kin}$ balancing gravitational
energy $E_{\rm grav}$:
\begin{equation}
{\textstyle\frac{1}{2}} \ddot I =
\underbrace{2 E_{\rm kin}}_{\sim4000\,{\rm eV\,cm^{-3}}	\times V} -
\underbrace{|E_{\rm grav}|}_{> \sim 1000\,{\rm eV\,cm^{-3}} \times V} +
\underbrace{3 P_{\rm int}}_{\sim10\,\rm eV\,cm^{-3}}	\!\!\!\!\!\!\!V \ -
\underbrace{3 P_{\rm ext}}_{<1\,\rm eV\,cm^{-3}}	\!\!\!\!\!\!V,
\end{equation}
where $I$ is the moment of inertia inside the considered volume $V$.  The
inner pressure $P_{\rm int}$ due to all the ISM components (gas, cosmic rays,
etc.) is a much too small energy reservoir to play any important role in the
virial equilibrium.  The outer pressure $P_{\rm ext}$ due to intergalactic gas
and radiation is even more negligible.  The only known large enough
{\it negative\/}
contribution to the virial balance susceptible to compensate the large
positive contribution of $E_{\rm kin}$ is gravitational, although the detected
mass is still insufficient.  Thus dark matter must
be invoked, particularly in the outer disks of spirals.

Now the equilibrium is certainly imperfect as we have seen above with the
large visible and IR fluxes ($\sim 10^{44}\,\rm erg\,s^{-1}$) through the
interstellar medium.  The quasi-static evolution of systems in near virial
equilibrium is to first order:
\begin{equation}
{\ddot I(t)} = \underbrace{{\ddot I(0)}}_{\sim 0} +
 t\,\frac{d {\ddot I(0)}}{dt} + {\cal O}(t^2)
\approx  2 t \left[ 2 \dot E_{\rm kin} \!+\!
\dot E_{\rm grav} \!+\! 3 \dot P_{\rm int} V
\!-\! 3 \dot P_{\rm ext} V \right] + {\cal O}(t^2) .
\end{equation}
Hence, while an equilibrium needs similar interacting energies, the slowest
quasi-static evolution needs also similar interacting {\it powers\/} cancelling
each others.  This is precisely what is suggested by FIR data on spirals.

Thus over time-scales of the order of $1-100\,\rm Myr$ a spiral may be
considered as in equilibrium at the largest scale because then the shortest
relevant evolution time-scale is the longer dynamical time.  However, at
smaller scale ($\sim 1\,\rm kpc$) the local power imbalance may be large due
to either the energy output by massive stars, or the fast radiative cooling
leading to dense molecular clouds.

Over time-scales of the order of $0.1-10\,\rm Gyr$ the heating resulting from
stellar activity and gas cooling in average must mostly cancel.  The feedback
mechanism regulating star formation at a constant {\it average\/} level via
dynamics is essential, otherwise one would expect the spiral as a whole either
to explode or to collapse rapidly.  A tight feedback requires a minimum galaxy
size in order to reduce the fluctuations; for too small galaxies such as
dwarfs, a couple of OB stars in excess is already significant to disrupt the
galaxy: there one may find a reason for a minimum size for star forming
systems looking like spirals\footnote{An upper spiral velocity weakly
dependent on their mass follows from the maximum time-scale $\tau_{\rm max}$
to consume all the nuclear energy in galaxies with the feedback $L \approx
v^5/G$. Then $\tau_{\rm max} \ll 0.008\, G M c^2 / v^5$, so $v <
(0.008\,GMc^2/12\,{\rm Gyr})^{1/5} \approx 500^{+300}_{-200}\,\rm km\,s^{-1}$
for $M=10^{11\pm1}\,M_\odot$.}.

Over longer time-scales, say $1-10\,\rm Gyr$, the gas consumption by a
sustained star formation is substantial, and eventually the feedback weakens.
Furthermore, other dynamical processes, such as a bar instability or mergers,
can become relevant and may modify the conditions of star formation.

After such considerations the initial conditions of galaxies look secondary
for the present state.  More important is the galactic micro-physics: star
formation and the ISM properties.  Physically this is a much more comfortable
situation because nowadays we expect rather chaotic initial conditions of
formation as more realistic than the ordered collapses envisioned for decades.
In order to obtain galaxies with systematic properties the information must be
encoded, as for stars, within the matter instead of the initial conditions.

%%%%%%%%%%%%%%%%%%%%%%%%%%%%%%%%%%%%%%%%%%%%%%%%%%%%%%%%%%%
\section{Stellar Dynamics as a First Order Tool}\label{StDy}
Consequently, stellar dynamics seems now clearly incomplete for the
understanding of spirals over $10-15\,\rm Gyr$. However, for shorter
time-scales like a few rotational periods, since the dominating energies are
kinetic and gravitational, we can indeed approximate the full Boltzmann
equation (as applied either to stars or molecules)
\begin{equation}
\frac{\partial f}{\partial t} + \vec v \cdot \vec\nabla f
- \vec\nabla \Phi \cdot\frac{\partial f}{\partial\vec v}  =
\left(\frac{\partial f}{\partial t}\right)_{\rm coll},
\end{equation}
by neglecting the right-hand side (rhs) collisional term.  {\it Stellar
dynamics can thus still be applied, but over time-scales much shorter than
believed earlier}.

For the optical parts of galaxies where the stellar mass dominates, the
mean-free path of stars is large and furthermore often collective effects can
be neglected because the kinetic energy in non-systematic motion is large
enough to yield not too small Jeans' lengths.  In sufficiently hot systems
collective effects are small at scale smaller than the Jeans' length, thus the
potential is replaced by a mean gravitational potential $\Phi$.  In
such a mean-field approximation galactic dynamics reduces to describing the
possible orbits in the potential $\Phi$,
\begin{equation}
\ddot{\vec x} = - \vec\nabla \Phi(\vec x).
\label{Motion}
\end{equation}
This is a problem belonging to the well studied Hamiltonian mechanics.  Let us
recall a few elementary general properties of such systems (e.g.~Arnold 1989),
useful to know before analyzing the phase space structure of galactic
potentials.

The first important concept is the one of {\bf orbit}, to distinguish from the
one of trajectory.  An orbit is the whole subset of phase-space generated over
an {\it infinite time\/} in the past and in the future by an initial condition
in a dynamical system such as Eq.(\ref{Motion}).  A {\bf trajectory} is the
subset of phase-space visited over a {\it finite\/} time by some initial point.
By construction an orbit is an {\it invariant\/} subset of phase space by the
motion, because all the points along an orbits generate trajectories within
it.  Orbits are thus the most fundamental invariant blocks with which we can
build equilibrium models of stellar systems.  Stellar systems may be seen as
made of orbits instead of stars.

There are several different types of orbits in 3D stellar systems which are
characterized by different {\it dimensions\/} $d$ in the 6-dimensional phase
space.  Either they have $d=0$ if they don't move, they are fixed points, or
at least $d=1$ if they move, and at most $d=5$ in equilibrium systems
due to the energy integral. The only orbits with
non-vanishing phase space volumes are usually the quasi-periodic orbits with
$d=3$ and the chaotic orbits with fractal, non-integer $d$.  The neighborhood
of fixed points and periodic orbits is either stable or unstable depending on
whether or not most of the neighboring trajectories remain close.  In case of
instability the neighborhood is mostly made of chaotic orbits.  These
properties are summarized as follow:
\begin{center}
\small
\begin{tabular}{lccc}
\tableline
\bf Orbit type	& \bf Dimension $d$ & \bf Neighborhood	& \bf Neighborhood \\
       		&		    & \bf if stable	& \bf  if unstable \\
\tableline
fixed point 	& 0 		    & periodic orbits	& chaotic orbits \\
periodic	& 1		    & quasi-periodic orb. & chaotic orbits \\
quasi-periodic	& 2, or 3	    & quasi-periodic orb. & --- \\
chaotic		& fractal: $1<d \leq 5$ & ---		& chaotic orbits \\
\tableline
\end{tabular}
\end{center}
The chaotic orbit class is the class of the orbits not classifiable in the
other
classes (like the `peculiar' galaxies in the Hubble classification).  It
includes different types of orbits which could be further sub-classified
according to their multifractal spectrum and fractal dimension.

As understood long ago by Poincar\'e, periodic orbits are certainly, after the
few fixed points, the most important class of orbits, not because many stars
are on them, but because when stable {\it they summarize the surrounding
phase-space}.  The stable periodic orbits are always surrounded by concentric
quasi-periodic orbits, or tori.  Those tori are much more numerous.  This
structure of concentric tori surrounding the periodic orbits in Hamiltonian
systems is the fundamental property allowing to conceptualize phase-space by
its stable periodic orbits.  This property is still insufficiently appreciated,
even by some dynamicists, motivating these elementary reminders.

Among the robust properties of orbits we note: 1) the shapes of the main
periodic orbits depend only on the potential {\it symmetries}, therefore we do
not need a precise description of the galaxy to know its orbital structure, 2)
the periodic orbits are insensitive to {\it low\/} spatial frequency
perturbations, so moderate low frequency symmetry breaking (such as a bar
bending out of the plane) do not change much the orbital structure, it only
deforms it slightly, and 3) the stable periodic orbits become {\it
attractors\/} by a weak dissipative perturbation.  A small dissipation in fact
condenses the trajectories of the system toward its most fundamental periodic
orbits.

However, periodic orbits are fragile against {\it high\/} spatial frequency
perturbations anywhere along their path.  For example, a small accumulation of
mass just at the galaxy center may profoundly change the stability of the
radial periodic orbits crossing the center, so the structure of the
surrounding phase space.

If we consider now all the neglected effects that should be taken into account
to reflect all the complexity of a real spiral, such as gas dynamics and star
formation, among all the possible structures of a pure stellar dynamical
representation {\it the main stable periodic orbits are certainly the ones that
survive the longest and are the best suited to describe evolution features}.
They become inevitable {\it concepts\/} to understand complex non-linear
dynamical processes.  This reality can be experienced by those dealing with the
complexity of $N$-body simulations, also including gas.  It is quite obvious
that, for example, when we think about a rotating disk we make a ``thought
economy'' by first considering only its circular orbits, even if no stars
follow exactly such orbits.

%%%%%%%%%%%%%%%%%%%%%%%%%%%%%%
\section{Phase-Space Structure of Barred Galaxies}\label{PhSpStructure}
The main dynamical feature of a barred galaxy is shaped by its only general
isolating integral, the Jacobi integral, also called the ``energy'' in a
rotating frame of reference.  The Hamiltonian in the rotating frame of the bar
reads
\begin{equation}
H = {\textstyle\frac{1}{2}}
(p_x^2+p_y^2+p_z^2) + \Phi(x,y,z) - \Omega_{\rm p}(x p_y - y p_x),
\end{equation}
where the momenta $p_x=\dot x-\Omega_{\rm p}y$, $p_y=\dot y+\Omega_{\rm p}x$,
and $p_z=\dot z$ are the velocity components in the instantaneous parallel
inertial frame.  The zero-velocity surface (ZVS) in the rotating frame
\begin{equation}
H_0 = \Phi(x,y,z) - {\textstyle\frac{1}{2}}\Omega_{\rm p}^2(x^2+y^2),
\end{equation}
bounds motion in space at low ``energy'' inside a football shape (nearly like
a spheroid) inside corotation.   Higher ``energy'' particles inside the bar
can escape first through tunnels in the ZVS near the end of the bar, and then
through the whole corotation circle at still higher ``energy''.  Four of the
five fixed points, the Lagrangian points, are located at the corotation.

The closest simple analytical model of a barred galaxy is an axisymmetric
disk, in which the basic unperturbed orbits are the circular orbits in the
principal plane.  The three orbital frequencies in the plane are the rotation
frequency $\Omega$, the radial and vertical epicyclic frequencies $\kappa$ and
$\nu$.  These frequencies squared depend only on the potential local
derivatives at $z=0$ (see Binney \& Tremaine 1987, p.~121).  Circular orbits
are then stable if $\Omega^2>0$, $\kappa^2>0$, and $\nu^2>0$.

If we consider the bar as a perturbation of the circular case, the circular
orbits are perturbed by the non-axisymmetric potential rotating at the
frequency $\Omega_{\rm p}$.  The resulting transverse deviations $\xi$ from
the circular orbits (either radial or vertical) can be described by the Hill's
equation, thats is the equation of a harmonic oscillator the natural frequency
$\omega_0$ of which is periodically modulated:
\begin{equation}
\ddot \xi + \omega_0^2\left[1 + \epsilon (t)\right] \xi = 0,
%\qquad \xi = \delta R, \ \ \omega_0 = \kappa, \qquad \mbox{\rm or}
%\qquad \xi = \delta z, \ \ \omega_0 = \nu,
\label{Hill}
\end{equation}
with $\epsilon(t)$ a small periodic function of frequency $\omega_{\rm per}$.
A general theorem (see e.g.~Arnold 1989) gives the {\it parametric
resonances\/}
conditions for arbitrarily {\it small\/} modulation $\epsilon(t)$:
\begin{equation}
\frac{\omega_0}{\omega_{\rm per}} =
\pm \frac{k}{2}, \quad k=0, 1, 2,\ldots\infty.
\end{equation}
With this theorem we can derive elegantly the resonance conditions, because we
do not need to know the precise form $\epsilon$ of the bar perturbation, but
only the perturbing frequency $\omega_{\rm per}$.   To determine the
resonance widths one does need however to know the particular form of
$\epsilon$.

The radial and vertical epicyclic frequencies $\kappa$ and $\nu$ are the
natural oscillations frequencies around the circular orbits, and the bar
perturbing frequency is $2(\Omega \mp \Omega_{\rm p})$ for direct/retrograde
circular orbits.  The factor 2 comes from the bi-symmetry of the bar (the
``number of arms").  Then we obtain the classical resonances when
\begin{equation}
\frac{\kappa }{ \Omega \mp \Omega_{\rm p}} = \pm m , \qquad
\frac{\nu    }{ \Omega \mp \Omega_{\rm p}} = \pm n ,
\quad \qquad m,n = 0, 1, 2,\ldots\infty.
\end{equation}
The main direct resonances encountered in barred galaxies occur for
$m=n=\infty$ (corotation), $m=n=2$ (radial and vertical inner Lindblad
resonances), $m=-2$ (radial outer Lindblad resonance), and $m=n=4$ (radial and
vertical ultra-harmonic resonances).  Around corotation we have the piling up
of an infinity of higher order resonances.  The main purpose of recalling
these simple considerations is to stress that the treatments of radial and
vertical resonances are the same, there is no ground to neglect the vertical
resonances, particularly because bars extend in the central regions of the
galaxies which are not much thinner than large.

In addition, the knowledge of the resonances gives the approximate {\it
shapes\/} of the nearby orbits.  For having elongated oval orbits we always
expect that the conditions are close to have a 2/1 Lindblad resonance, because
an oval is a 2/1 perturbation of a circle.  Rectangular shapes are associated
with 4/1 resonances, etc.  This applies in the vertical direction too, a 2/1
resonance is associated with 2/1 vertical oscillations out of the plane, etc.
Once the typical resonances in barred galaxies are known, it is simple to
guess how the main periodic orbits look like.  Detailed numerical calculations
of orbits can help in quantifying precisely the orbit positions and shapes,
and also in finding higher order periodic orbits.  Numerous works have been
made in this area allowing to grasp the main general features of the bar
dynamics (see references, e.g., in the review of Sellwood \& Wilkinson 1993).
The general properties are well understood in a descriptive way by several
numerical calculation works.  Although the main periodic orbits families in the
plane and in 3D, including their shape, are well known by now, there is still
no simple but realistic analytical model of galactic bars.

Exactly similar considerations to circular orbits apply for radial periodic
orbits. A slight complication is that for the $z$-axis orbits in a rotating bar
the two transverse frequencies are coupled by rotation, which leads to a
generalizations of the Hill's equation and new possibilities of parametric
resonances via ``complex instability'' (Pfenniger 1987).  But in non-rotating
bars, Hill's equation (\ref{Hill}) applies for each transverse component
of the three types of axial orbits, the perturbing frequency being the
oscillation frequency of the radial orbit.  Then a 2/1 resonance leads to
banana shaped orbits (shaping the neighboring phase space) which favors a
bending instability in non-rotating ellipsoidal systems too.

For oblate axisymmetric disks with positive density (constraining the
frequencies by $2\Omega^2\! <\! \kappa^2+\nu^2$, and $\Omega\! =\! c\nu$,
where $c\!<\!1$ is the potential axis ratio) a central mass concentration {\it
always\/} produces low order radial {\it and vertical\/} resonances.  Now as
long as the potential remains axisymmetric the width of the resonance is zero,
and these resonances are ineffective.  But as soon as a triaxial deformation
exists, the resonance widths grow which fosters chaotic motion.  Then
diffusion in the vertical direction is natural near the center of a galaxy,
especially if slow dissipative processes accrete mass near the bottom of the
potential.  Then the idea that a bulge may naturally grow in a barred
potential is dynamically justified (Pfenniger 1984, 1985; Pfenniger \& Norman
1990).

The dissolution of a bar by a growing central mass concentration is an example
of possible qualitative prediction allowed by knowing the periodic orbits of
bars (Hasan \& Norman 1990; Pfenniger \& Norman 1990; Hasan et al.~1993).
Since the radial and vertical ILR radii increase rapidly when the central mass
concentration increases $R_{\rm ILR} \propto M_{\rm cm}^{2.8}$, and the
elongated orbits supporting the bar are replaced by orbits perpendicular to the
bar inside the growing ILR radius, the elongated shape of the bar is rapidly no
longer compatible with the growing central mass concentration.  Orbit
calculations show that around $5-15\%$ percents of the disk mass as a central
mass concentration should be sufficient to destroy the bar.  Since the vertical
ILR is typically close to the radial ILR, one expects an extended 3D diffusion
of the initial central part of the disk.

To conclude this Section, simple periodic orbits considerations show that
barred galaxies have numerous resonances and associated chaotic zones from
which we can {\it expect\/} instabilities.  For studying these we need however
techniques able to take self-gravitation into account, such as $N$-body
methods.

%%%%%%%%%%%%%%%%%%%%%%%%%%%%%%
\section{Types of Instabilities in Collisionless Gravitating Systems}
\label{InCoGR}
Many works have considered the problem of self-consistent stability with the
linearized Collisionless Boltzmann equation.  Here we just point out how this
method may help in distinguishing two different sources of instabilities in
real stellar systems, simultaneously clarifying the limitations of the method.

Suppose we know a solution $f_0$, $\Phi_0$ ($\rho_0=\int{f_0 \,d^3v}$) of the
Collisionless Boltzmann and Poisson equations.  Then, as usual, we want to
describe the evolution of small perturbations $\delta f$, $\delta\Phi$ around
the assumed known solution $f_0$, $\Phi_0$ in the linear approximation.
Linearizing the equations we obtain:
\begin{eqnarray}
\label{linCB}
& \overbrace{\frac{\partial \delta f}{\partial t} +
    \vec v \cdot \vec\nabla\delta f-\vec\nabla\Phi_0 \cdot
    \frac{\partial \delta f}{\partial \vec v}}^{
    \mbox{\small orbital behavior, mixing of $\delta f$}}
\ \equiv\
{\displaystyle{\frac{{\rm D} \delta f }{{\rm D} t}} }
\quad=\quad
\overbrace{\vec\nabla\delta\Phi \cdot
	\frac{\partial f_0}{\partial \vec v}}^{
	\mbox{\small source of $\delta f$}} & ,\\[1mm]
& \vec\nabla\cdot\vec\nabla\delta\Phi \ =\  4\pi G\delta\rho &.
\end{eqnarray}
As set, Eq.(\ref{linCB}) shows that the variations $\delta f$ depend on two
simple effects.

First, the rhs of Eq.(\ref{linCB}) shows that to modify the density of a volume
element moving around a given unperturbed orbit, a gradient of $f_0$ in the
velocity space is required.  Furthermore, only the force fluctuation component
parallel to the velocity gradient contributes.  Large density gradients in
velocity space favor instabilities parallel to the gradient (so radial in cold
rotating disks).

The second factor of effective instability is a little more subtle.  The
left-hand side (lhs) term, $D\delta f/D t$, describes the Lagrangian derivative
in phase space of a fluctuating element along an unperturbed orbit. So, even if
the rhs vanishes, or is very small, the {\it form\/} of a compact element may
be rapidly modified into a threaded structure the fineness of which becomes
smaller than allowed by the various approximations.  This occurs precisely
around unstable orbits in the unperturbed system, such as chaotic orbits.  This
sensitive dependence means that the neglected terms (higher order terms,
collision and dissipative terms) become rapidly relevant for the real
system\footnote{The same occurs in many other problems: when the linear
operator $A$ of the problem $Ax = B$ has widely different {\it singular
values\/} then the problem is {\it ill-posed\/} and the solution $x$ is
ill-defined.}.

Thus in stellar systems which contain chaotic regions of phase space, so
typically around resonances, an initial compact volume element of $\delta f$
is rapidly distorted into very intermingled shapes, the slightest perturbation
is exponentially amplified, and non-linear considerations are very soon
necessary.  In such a situation it is more natural to ascribe the resulting
instability not to the rhs source term, but to the lhs phase space structure.
On the other hand, when the orbital motion is regular, a compact initial
volume element remains compact much longer, the lhs is well-behaved and a
growing mode can only be attributed to the rhs term.  In such cases
the instability can be ascribed to the full distribution function
$f_0$.  Note that analytical models have usually a regular phase space, so the
problem appears preferentially in numerical models not biased against chaos.

In summary, in order to progress in the understanding of stellar systems and
their instabilities we can divide their analysis in two distinct parts:
\begin{enumerate}
\item
The understanding of the phase-space structure (i.e.~the orbits) of the
potential $\Phi_0$ independently of the velocity structure of $f_0$.
The resonance
regions and the chaotic zones are then susceptible to seed collective
instabilities and non-linear effects not describable with a linear theory.
\item
The understanding of the specific self-gravity effects associated with
particular $f_0$'s consistent with a given $\Phi_0$.  One may generally expect
that sufficiently cold distribution functions should produce instabilities
not directly related to the previous aspect.  For a regular phase-space the
linear mode analysis should work much better.
\end{enumerate}
As illustration, in spirals the bar instability comes mainly from a too cold
$f_0$, because phase space surrounding stable circular orbits is regular,
while the bending instability leading to peanut-shaped bars, discussed below,
is mainly due to a 2/1 vertical resonance and is little dependent on the
velocity structure of $f_0$.

\section{Smoothness Assumption}
An assumption which is most of the time not discussed is that $f$ should be a
differentiable function.  In ordinary gases with molecules having short range
interactions and frequent collisions, any irregular distribution is rapidly
smoothed out locally by the ``molecular chaos''.  The smoothing principle
comes then from the short relaxation time and the lack of long-range
interactions, which allows a fast local decorrelation and homogenization of
the particles.

In collisionless gravitating systems this rapid smoothing is far from being
obvious and demonstrated.  As soon as instabilities occur, the long range of
gravitation correlates the fluctuating part $\delta f$ much more than in
ordinary gas, and since we lack of a smoothing principle, no good reason other
than commodity allows to assume that a differentiable $\delta f$ is a valid
assumption allowing its use in the variational differential equation.

In fact numerical experiments indicate rapid limitations of the linear theory.
For example Toomre \& Kalnajs (1991) have simulated a small portion of a
self-gravitating flat disk which is maintained slightly unstable.  The results
show clearly that long range correlations are rapidly created and do persist;
a fractal scale invariant and dynamical state follows.  The linearization
approach is of no use for describing such states.  Another well documented
case of gravitational instability occurs in the numerous simulations of
expanding universes.  The runs develop usually non-linear fractal structures
which cannot be derived from a linear study.  In such a situation, analogous
to turbulence in fluids, the system has a sensitive dependence on small-scale
effects and perturbations.  The collisional as well as weak dissipative
effects can be crucial. Therefore we should be cautious about drawing
conclusions without a thorough non-linear analysis.  $N$-body methods are
presently essential to study non-linearities with the advantage that they
include {\it non-vanishing collisional effects}.

%%%%%%%%%%%%%%%%%%%%%%%%%%%%%%
\section{Self-Consistent 3D Barred Galaxy Models}
In the early 80's it was generalized from a single peculiar edge-on S0 galaxy
with a small(!) bulge (NGC\,4762) that bars are generally flat ($a/c\sim10$)
(see Kormendy 1982).  Theory says such flat bars are implausible to maintain
for a long time since they imply highly anisotropic velocity distribution, a
strong velocity gradient of $f_0$, and strong vertical resonances from the bar
potential.

In fact in 3D $N$-body simulations most of the initially flat bars thicken
rapidly and are subject to bending instabilities transverse to the plane
(Friedli \& Pfenniger 1990; Pfenniger \& Friedli 1991 (PF91); Raha et
al.~1991).  The end-result of these bending instabilities are box- or
peanut-shaped bars (Miller \& Smith 1979): stable structures over several Gyr
resembling much the observed peanut-shaped bulges according to viewing angle
of the bar (Combes \& Sanders 1981; Combes et al.~1990; see also references in
Merrifield, this volume).  The peanut-shaped bars look round when viewed
end-on.

While Raha et al.~have identified this instability with the ``fire-hose
instability'' (the out of the plane instability of an infinite homogeneous
thin sheet), the orbital analysis in PF91 shows clearly that the fire-hose
instability picture is a too rough analogy to predict the bending instability
main characteristics such as its size, the principal mode of bending, and the
instability threshold.  The detailed analysis of the $N$-body bar run in PF91
(see also Pfenniger 1990) greatly improves the understanding of the
instability by considering not only 1) its simple morphological and kinematic
description, but also 2) its potential resonances, which give a first idea on
the orbits, further 3) its periodic and other orbits at different times, and
then 4) its distribution function changes.  Finally the whole non-linear
evolution of the ensemble can be much better understood.

Non-linearities are essential all along the phenomenon: the instability starts
near the vertical 2/1 resonance, so is associated with this resonance, it
chooses to bend up or down randomly from the fluctuations, with precisely a
2/1 banana shaped mode, then it symmetrizes rapidly its vertical profile. The
instability saturates around 2/2/1 stable periodic orbits, which pre-exist,
accompany and survive the instability, explaining the final peanut shape.  As
the instability proceeds, the vertical 2/1 resonance sweeps the bar
particles from low to high ``energy'', allowing most of them to leave the
plane up to heights and distances allowed by the ZVS.  Thus, the bending
instability is a nice example of a gravitational instability {\it little
dependent on the velocity space structure of $f_0$}, it occurs also in bars
initially far from being flat.  Clearly, pure orbital considerations, lacking
of the self-gravity, are insufficient to predict firmly the bending and time
evolution.  But a linear mode analysis such as in Merritt \& Sellwood (1994)
is unable to describe the whole phenomenon too\footnote{In order to explain
this surprisingly fierce instability, these authors must finally distinguish
it from the fire-hose case and rely on giving an explanation in term of the
``oscillations of stars'' with a 2/1 frequency ratio. Of course the orbit
description is just the same in a systematic fashion.  The reported strong
grid dependence in their numerical results seems natural once realized that
around a major resonance chaos causes a sensitive dependence not only on the
neglected non-linear terms, but also on the numerical technique; this is just
the signature of a so-called ``ill-posed'' problem mentioned above.}.  It is
only the association of the two approaches together with the $N$-body
technique which allows a detailed understanding of the entire process.

As noted in Section \ref{PhSpStructure}, the knowledge of the orbital
structure in bars is also useful for predicting conditions of their
destruction.  This can be tested with $N$-body runs, which confirm that indeed
a mass accumulation within the original ILR changes the orbital structure to
an extent that elongated bars are no longer possible, but only 1-3\% of
additional mass is required.  Such an instability involves either a
dissipative factor in order to grow a central mass concentration (Friedli \&
Pfenniger 1991; Sellwood, this volume), or dynamical friction of dense
satellites (Pfenniger 1991, 1992, 1993).  In this respect an accretion of
5-10\% of mass of satellites inside the bar region is able to destroy the bar
and form a much bigger bulge like the one in M104.

Other indications from $N$-body simulations that the dynamical picture that we
propose is essentially correct, are 1) the reshaping of an initial disk into
an exponential disk plus a steeper bulge-like profile in the bar region
results automatically from the dynamical effects of a bar (Hohl 1971), 2) in
such conditions the velocity ellipsoid tends to be anisotropic with an
exponential profile consistent with the Milky-Way observations (Lewis \&
Freeman 1989): $\sigma_R^2 \propto \exp(-R/h)$, with $\sigma_R> \sigma_{\phi}
>\sigma_z$ (PF91).

All these different kinds of dynamical evolutions give a living character to
the spirals, contrasting with older static views.  It is then natural to assume
that bulges may grow secularly from their disk, which is consistent with the
observations that bulge stars are metal rich (Rich 1992).  The general picture
of disk dynamical evolution is a sequence of barred and unbarred phases through
which the bulge size grows irreversibly.  This is one of the several arguments
for proposing an evolution of the spirals along the Hubble sequence from Sd to
Sa (Pfenniger et al.~1994).  It is not clear yet how bars can be recurrent,
because a central mass concentration once formed should prevent further bar
formation unless a large amount of cold and angular momentum rich material is
able to accrete quietly on the galaxy disk without heating it.

Finally, a general observed trend concerns the apparent near integrability of
stable stellar systems.  The phase-space analysis of $N$-body runs, such as in
PF91, shows that the final most stable configurations are free of strong
chaotic orbits and have remarkably simple phase space structures reminiscent of
integrable systems, at least for the populated phase space regions.  The
remaining ubiquitous weak chaos is indistinguishable from the particle noise.
This is consistent with the stability considerations given in
Sect.~\ref{InCoGR}: chaotic regions are likely to seed collective
instabilities.

%%%%%%%%%%%%%%%%%%%%%%%%%%%%%%
\section{Final Remarks and Conclusions}\label{Conclusions}
The first order dynamics determining the shape and evolution of barred
galaxies is now well understood.  The component decomposition of these
galaxies in bulge, bar, and disk is dynamically illusory since many stars
constantly switch from one ``component'' to the next and back.  The gross
morphology of bars, their sense of evolution is well understood, and many
observed features like the exponential disks, the peanut-shaped bars, and the
SB0 profiles can be remarkably reproduced by numerical means.
Finer morphologies like rings, ansae, double bars are now at the limit of the
numerical resolution in fully self-consistent models.

For understanding galaxies the knowledge of the periodic orbits turns out to
be an invaluable tool to simplify the description of the dynamical processes,
and to develop an intuition allowing even correct guesses without computer!

Bars can form spontaneously from their disk, peanut-shaped bulges can grow
from their parent bar, and bars can fully dissolve in spheroidal bulges
through a central accumulation of mass, so secular dynamical evolution of
spirals appears as natural, with typical time-scales of the order of
$0.1-10\,\rm Gyr$ for changing significantly the spiral type.

\acknowledgments This work is supported by the Swiss FNRS.

%%%%%%%%%%%%%%%%%%%%%%%%%%%%%%%%%%%%%%%%%%%%%%%%%%%%%%%%%%%%%

\begin{question}{B. Elmegreen}
What does a bar look like after it is destroyed?  It seems impossible that
early type bars can form bulges that big and also that late type bars form
bulges, because late type galaxies don't have significant bulges.  Perhaps
only some S0 galaxies with the largest bulges contain totally destroyed bars.
\end{question}
\begin{answer}{D. Pfenniger}
In our simulations a fully destroyed bar looks like a spheroid with an
extension similar to the original bar.  Now I would interpret systems with
both bar and bulge as systems right in the process of dissolving the bar;
during this stage the bulge size extends up to the ILR, so can be much smaller
than the bar such as in late types.  In early systems with big bulges, like
M104, the size of the bulge is too big to result only from a single bar, from
energetics considerations.  But our quoted simulations do produce such systems
by merging satellites involving in total about $5-10\%$ of the stellar mass;
such merging events also destroy a pre-existing bar.
\end{answer}

\begin{question}{S. Dodonov}
In the 5-Gyr N-body bar the ``hot'' population showed two distinct bumps in
the angular momentum distribution. What are they?
\end{question}
\begin{answer}{D. Pfenniger}
The primary bump is made of the stars trapped by the bar.  The secondary bump
at higher angular momentum but confined inside corotation consists of
particles around the 4/1 and higher order resonances.
\end{answer}

\end{document}